\documentstyle[12pt]{article}

\newcommand{\be}{\begin{equation}}
\newcommand{\ee}{\end{equation}}
\newcommand{\bea}{\begin{eqnarray}}
\newcommand{\eea}{\end{eqnarray}}
\newcommand{\barr}{\begin{array}}
\newcommand{\earr}{\end{array}}

\def\be{\begin{equation}}
\def\ee{\end{equation}}

\def\ba{\begin{array}}
\def\ea{\end{array}}
\def\bea{\begin{eqnarray}}
\def\eea{\end{eqnarray}}

\def\gm{\gamma}
\def\rc{r_c}
\def\met{\eta_{\mu\nu}}
\def\vphi{\langle \phi \rangle}
\def\tphi{\tilde{\phi}}
\def\mphi{m_{\phi}}

\begin{document}

\hfill{MRI-PHY/P2000-05-34}

\hfill{PRL-TH/2000-26}

\bigskip

\centerline{\bf Effects of non-factorizable metric}
\centerline{\bf on neutrino
 oscillation inside supernova}
\vspace{1in}
\bigskip

\centerline{\bf Uma Mahanta}
\centerline{ Mehta Research Institute, Chhatnag Road, Jhusi, Allahabad-211019}
\centerline{ and}
\centerline{\bf Subhendra Mohanty}
\centerline{Theory Group, Physical Research Laboratory,
 Navarangpura, Ahmedabad}

\vspace{1in}
\bigskip
\centerline{\bf Abstract}
\bigskip

In this paper we construct the interaction potential between the dense
supernova core and neutrinos due to the exchange of light radions in the
Randall-Sundrum scenario. We then show that the radion exchange potential
affects the neutrino oscillation phenomenology inside the supernova
significantly if the radion mass is less than around 1 GeV. In order 
that the Bethe-Wilson mechanism for heating the envelope and
r-process neucleosynthesis be operative, the
radion mass must be greater than 1 GeV. Bounds on the radion
mass of the same order of magnitude can also be derived from
TASSO and CLEO data on B decays.

\vfill\eject

\centerline{\bf I. Introduction}

\medskip

Several radical proposals based on higher dimensional spacetime have been
put forward recently to explain the large hierarchy between the Planck
scale and the electroweak (EW) scale. Among them the Randall-Sundrum (RS)
scenario [1] is particularly attractive since it proposes a higher dimensional
spacetime with a non-factorizable five dimensional metric of the form
\be
ds^2= e^{-2k \rc
\vert \theta\vert }\met dx^{\mu}dx^{\nu}+\rc^2 d\theta^2
\ee
where k is a scale of the order of the Planck mass, $x^{\mu}$ are the 
Lorentz coordinates of four dimensional surfaces of constant $\theta$.
$-\pi\le\theta\le \pi$ is the angular coordinate of the extra dimension
which is an $S^1\over Z_2$ space with orbifold symmetry. The points
(x, $\theta $) and (x, -$\theta $ ) are therefore identified. 
The compactification radius $\rc$ is the vacuum expectation
value (vev) of the modulus field $T(x)$. 
 Two 3-branes, the hidden and visible branes extending
in the $x^{\mu}$ directions are located at the orbifold fixed points
0 and $\pi$. The existence of the modulus field
is therefore the most direct and straightforward result of non-factorizable
metric. In other words the effects of the nonfactorizable metric on the low 
energy phenomenology of the visible brane can be described in terms of the
couplings of the modulus field to the SM fields. Randall and Sundrum showed
[1] that the particular nonfactorizable metric of eqn (1) satisfies the five
dimensional Einstein equations for this set up if the vacuum energies
of the two 3-branes and the bulk cosmological constant are related
in a particular way through the single scale factor k.

In the original proposal of Randall and Sundrum the modulus field was
massless and had zero potential. The vev of the modulus was therefore
not stabilized. Clearly a massless radion with TeV scale couplings
is phenomenologically unacceptable since it would lead to a long
range universal force that is 32 orders of magnitude stronger than
gravity.
 Goldberger and Wise [2] showed that the modulus
of RS scenario  can be stabilized by introducing a scalar field in the bulk
with suitable interaction potentials on the two 3-branes.
 In this work
we shall assume that the modulus is stabilized by Goldberger-Wise
mechanism and has a non-zero mass.
 By considering
fluctuations of the radion field $\phi$ (which is related to the
modulus field T(x) through the eqn $\phi (x) = f e^{-k\pi T (x)}  $ with 
$f\approx M_{pl}$) about its vev it was shown in [3] that the radion couples
to the visible brane matter through the trace of the
energy momentum tensor
\be
L_I= T^{\mu}_{\mu}{\tphi\over \vphi}
\ee
Here $T^{\mu}_{\mu}$ is the trace of the energy momentum tensor of the
of visible brane matter. $\tphi =\phi - \vphi $ is the fluctuation 
of the radion field
from its vev $\vphi$. In the RS scenario $\vphi $ is in the 
 TeV range for
$k\rc \approx 12$ which is needed in order to generate the weak scale from
the Planck scale through the exponential warp factor.
Note that the radion coupling to matter fields on the visible brane given
by eqn (2) is universal since it arises from the non-factorizable metric.
The radion therefore always couples to the trace of the energy momentum 
tensor of the relevant fields or effective degrees of freedom on the visible
brane.
 Inside the
supernova the relevant matter fields are the nucleons and the radion 
coupling to them is given by 
\be
L_{\phi N}={\langle (T^{\alpha}_{\alpha})_m\rangle\over  \vphi}
 \tphi\approx {\rho_m\over \vphi}\tphi
\ee
Here we have treated the nucleons inside the supernova as non relativistic.
$\rho_m$ is the matter density for non relativistic supernova matter.
On the contrary the higgs boson has tree level couplings only to
SM fermions and weak gauge bosons. The higgs coupling
 to nucleons arises from two gluon
intermediate state which are emitted from the vertices of a triangular 
top quark loop. The higgs boson is connected to the third vertex. Effectively
one integrates over virtual heavy quark fluctuations (which occur at distance
scales of ${1\over m_Q}\ll $ 1 fm) to get the higgs coupling to nucleons.
The usual loop suppression factor makes the higgs coupling to
nucleons much smaller than that of the radion coupling (${m_N\over \vphi}$).
 We shall therefore
ignore the effect of higgs exchange between the supernova matter and 
neutrinos in comparision to the radion exchange potential.
The radion coupling to SM fermions on the visible brane is given by
\be
L_{\phi f}={m_f\over \vphi}\bar {f}f \tphi 
\ee
Neutrinos  can however have both Dirac and Majorana mass.
But the neutrino oscillation probability at high energies
is the same for either type of neutrino mass. In this paper
we shall assume for definiteness
that the radion couples to neutrinos through their
Dirac mass term.

\medskip

\centerline{\bf II. The radion exchange potential for 
neutrinos inside supernova}

\medskip

In this section we shall construct the radion exchange potential for 
neutrinos inside the supernova. Consider the scattering amplitude due
to radion exchange between nonrelativistic supernova matter and neutrinos
\be
M= {1\over \vphi ^2}\langle (T^{\alpha}_{\alpha})_m\rangle {1\over
k^2-\mphi^2}\langle  (T^{\beta}_{\beta})_{\nu}\rangle 
\ee
where $\mphi$ is the radion mass. The trace of the stress tensor
for non relativistic supernova matter is given by 
$\langle (T^{\alpha}_{\alpha})_{mat})\rangle =\rho_m$. 
 The trace of the
energy momentum tensor for neutrinos is given by 
$\langle 
(T^{\beta}_{\beta})_{\nu}\rangle =m_{\nu}\bar \nu (p)\nu (p)$. Here
we shall assume forward scattering of  neutrinos i.e $k\ll p$.

The exchange of radions between the supernova matter and neutrinos therefore
generates an interaction potential for neutrinos which is given by
\be
V_{\nu}=-{1\over \vphi ^2 \mphi^2} m_{\nu} \rho_m
\ee
Unlike the weak interaction potential the radion exchange
potential (6) has the same sign for neutrinos and antineutrinos.

\medskip 

\centerline{\bf III. Effect of radion exchange potential on $\nu$
oscillation in supernova}

\medskip  

Neutrino oscillations between two different mass eigenstates
$\nu_1$ and $\nu_2$ is given by the equation
\be
i {d \over dt } \left( \barr{c} \nu_{1 } \\ \nu_{2} \earr \right)
= \left( ~~\left(\barr{cc} 0 & 0\\ 
0  & {\Delta m^2 \over 2 E} - {\Delta m^2\over 2 m_\nu}
 {~\rho_{m}\over  m_{\phi}^2~\langle \phi \rangle^2}
\earr \right) + H_{weak} \right)
\left( \barr{c} \nu_{1 } \\ \nu_{2} \earr \right)
\label{h}
\ee
where $\Delta m^2 = (m_2^2 -m_1^2)$.
The weak interaction potential in the flavour basis (say $\nu_e,\nu_\mu$ ) is
\bea
H_{weak}\left( \barr{c} \nu_{e } \\ \nu_{\mu} \earr \right) = 
\left(\barr{cc} \sqrt{2} G_F N_e & 0\\
0  & 0
\earr \right) \left( \barr{c} \nu_{e } \\ \nu_{\mu} \earr \right)
\label{hm}
\eea

We can write (7-8) in the flavour basis as
\bea
i {d \over dt } \left( \barr{c} \nu_{e } \\ \nu_{\mu} \earr \right)
= \left(\barr{cc} 0 & b
\\b   & a \earr \right) 
\left( \barr{c} \nu_{e } \\ \nu_{\mu} \earr \right)
\eea
where
\be
a= \left({\Delta m^2 \over 2 E} - {\Delta m^2 \over 2m_\nu}
{~\rho_{m}\over 2 m_{\phi}^2 \langle \phi \rangle^2}\right) cos 2 \theta
- \sqrt{2} G_F N_e
\label{a}
\ee
and 
\be
b=\left({\Delta m^2 \over 2 E} - {\Delta m^2 \over 2 m_\nu}
{\rho_{m}\over 2 m_{\phi}^2~ \langle \phi \rangle^2}\right)~
{1\over 2} sin 2\theta\nonumber\\
\label{b}
\ee
 where $\theta$ is the vaccum mixing
angle
between $\nu_e$ and $\nu_\mu$ and $N_e$ is the number density of
electrons.
One can see from (10, 11) that the radion exchange potential will
 be significant in neutrino oscillation if
\be
 {\Delta m^2 \over 2 E} \leq {\Delta m^2 \over 2m_\nu}
{~\rho_{m}\over 2 m_{\phi}^2 \langle \phi \rangle^2}
\ee
i.e when the radion is light and satisfies the following inequality
\be
m_{\phi}^2 \leq {~\rho_{m}\over   \langle \phi \rangle^2}{E_\nu \over m_\nu}
\label{mphi}
\ee
The most stringent bounds on $m_\phi$ can therefore be obtained by
considering situations where the neutrino energy and the matter density
are large. These criteria are ideally met by considering neutrino
oscillations in supernovae where matter density $\rho_{m} \sim 10^{14}
gm/cm^3 $. The neutrino energies are in the range of
$E_\nu = 50 MeV$ . Taking $ m_\nu \leq 1$ eV  we obtain the bound
$m_\phi > 1 $ GeV  in order that the neutrino oscillations
in supernovae be unaffected by the moduli exchange potential. 
In the supernovae the normal MSW resonance [4,5] due to standard
weak interactions  occurs in a cosmologicaly
interesting mass range of $\Delta m^2 \sim 1 $ eV. 
If the moduli mass is less than 1 GeV then the resonant conversions
of electron neutrinos to other species will not occur as 
a  (in eqn. 10 ) will be negative and there will be  no resonance .

In supernovae neutrino oscillations are needed to explain the
transfer of energy to the outer envelopes by the Bethe-Wilson mechanism
[6], and also to explain the r-process nucleosynthesis
[7]. If the moduli mass is
smaller than a 1 GeV then these cannot take place within a
cosmologically acceptable mass range of the neutrinos. 

The neutrino oscillation induced by the Randall-Sundrum radion discussed
in this report is a matter induced effect like the well known MSW
mechanism. It vanishes if the neutrinos are degenrate but not massless.
In this respect it differs from the Damour-Polyakov (DP) dilaton
[8] which occurs in the in the effective theory of gravity derived 
from strings. Like the RS radion it also couples to the trace of the 
energy momentum tensor. The neutrino oscillation induced by the DP
dilaton does not vanish [9] if the neutrinos are completely degenerate
but not massless. Because of this property the neutrino oscillation
induced by the DP dilaton has been proposed as a possible solution
to the solar neutrino problem.
We would like to note that the radion coupling to visible brane matter
given by eqn(2) has been derived by neglecting the back reaction of the
bulk scalar (responsible for stabilizing the modulus) 
on the background metric. In this approximation the metric on the visible
brane is given by $({\phi\over f})^2 \eta_{\mu\nu}$.
If the backreaction of the bulk scalar on the background metric
 is taken into account the
resulting warp factor might depend on the radion field in a different
way. This might change the radion coupling to visible brane matter.
It would be interesting to examine the effects of such changes
on  the estimates presented in this report.
 
\medskip

\centerline{\bf IV. Comparing the supernova bound on $\mphi$
with bounds from B decays}

\medskip

In this section we shall present the bounds on light radion mass
between $2m_{\mu}$ and $2m_{\tau}$ that
 could be derived from B decays. The radion coupling to SM fermions
is similar to that of the Higgs coupling except for a scale factor of
${v\over \vphi}$. Here v is the electroweak symmetry breaking scale.
 Therefore these bounds can be derived from similar
bounds on $m_h$ obtained from B decays by inserting the appropriate scale 
factor. In the following we shall assume that $\vphi =$ 1 TeV.

Bounds on  very light radion can be obtained both from inclusive and 
exclusive B deacys. However in this work
we shall consider the bounds from inclusive B decays
only since the hadronic uncertainties in their theoretical estimates
are much less than those in exclusive decays. The most important inclusive
B deacy limits on $\mphi$ between 2$m_{\mu}$ and 2$m_{\tau}$ are as follows

1. Radion masses in the range $2m_{\mu}< m_{\phi}< 2m_{\pi}$ can be ruled
out by the TASSO limits [10] on Br($\tphi \rightarrow \mu^+\mu^- X$). In this
mass range it is assumed that
Br($\tphi\rightarrow \mu^+\mu^- )\approx $1 which is valid to a very good
 approximation.

2. It may be possible to use the TASSO limit to rule out $\mphi$ between
500 Mev and 1 Gev  [11] if the $\tphi\rightarrow \pi\pi$ enhancement does not
severly suppress the $\tphi\rightarrow \mu^+\mu^-$ branching ratio.
However radion masses in the vicinity of the scalar resonance
f$_0$(975 MeV)  and other resonances  
in this region cannot be ruled out due to strong $\pi\pi$ final 
state interactions.

3. For 1 GeV$< m_{\phi}< 2m_{\tau}$ the CLEO limits [12]
 can be used to rule out
existence of a radion so long as Br($\tphi\rightarrow \mu^+\mu^-)> 2.5
\times 10^{-3}$. The J/$\psi$ region is of course left out.

4. For $\mphi> 2m_{\tau}, 2m_c$ the dominant deacys are 
$\tphi\rightarrow \tau^+\tau^-$ and $c\bar{c}$ for which there are no 
reliable bounds.

We would like to note that the LEP and Tevatron bounds on radion mass
presented in refs [13] and [14] do not exclude radions with mass between
100 MeV and 1 GeV. The LEP and Tevatron bounds on $m_{\phi}$ presented in 
these reports were derived based on $\tphi\rightarrow b\bar {b}$
and $\tphi\rightarrow \gm\gm$ decay modes of the radion respectively.
However the branching ratios of these decay modes become negligible
for radions with mass less than 1 GeV. This emhasizes the importance of
excluding radions in the 100 MeV-1 GeV range by indirect methods like 
B decays and neutrino oscillation inside the supernova.

In conclusion in this report we have derived a lower bound on the 
radion mass so that radion exchange potential does not significantly
affects the usual neutrino oscillation phenomenology inside the
supernova. This bound is of the order of 1 GeV. Bounds on the radion
mass of the same order of magnitude can also be derived from TASSO
and CLEO limits on inclusive B decays.

\medskip

\centerline{\bf References}

\medskip

[1] L. Randall and R. Sundrum, Phys. Rev. Lett 83, 3370 (1999).

\medskip

[2] W. D. Goldberger and M. B. Wise, Phys. Rev. Lett. 83,
04922 (1999); Phys. Rev. D 60, 107505 (1999).

\medskip

[3] C. Csaki, M. Graesser, L. Randall and J. Terning, 
hep-ph/9911406; W. D. Goldberger and M. B. Wise, hep-ph/9911457.

\medskip

[4] L. Wolfenstein, Phys. ReV. D 17, 2369 (1978).

\medskip

[5] S. P. Mikheyev and A. Yu. Smirnov, Sov. J. Nucl. Phys. 42, 913
(1985).

\medskip

[6] H. A. Bethe and J. R. Wilson, Astrophys. J.295, 14 (1985).

\medskip

[7.] Y. Z. Qian et al, Phys. Rev.Lett. 58, 1965 (1993).

\medskip

[8.] T. Damour and A. Polyakov, Gen. Rel. Grav. 26, 1171 (1994);

Nucl. Phys. B 423, 532 (1994).

\medskip

[9.] A. Halprin and C. N. Leung, Phys. Lett. B 416, 361 (1998).

\medskip

[10.] A. Snyder et al., Phys. Lett. B 229, 169 (1989).

\medskip

[11] M. Althoff et al., Z Phys C 22, 219 (1984).

\medskip

[12.] P. Avery et al., Phys. Rev. Lett 53, 1309, (1984).

\medskip

[13.] U. Mahanta and S. Rakshit, hep-ph/0002049.

\medskip

[14.] U. Mahanta and A. Datta, hep-ph/0002183.

\end{document}